\documentclass[a4paper]{report}
\usepackage[utf8]{inputenc}
\usepackage[T1]{fontenc}
\usepackage{RJournal}
\usepackage{amsmath,amssymb,array}
\usepackage{booktabs}

\usepackage{algpseudocode}
\usepackage{algorithm2e}
\usepackage{hyperref}
\usepackage{bbm}
\usepackage{amsmath}
\RestyleAlgo{ruled} 
\usepackage{floatrow}
\usepackage{textgreek}
\usepackage{blindtext}
\usepackage{scalerel}
\newcommand\Tau{\scalerel*{\tau}{T}}
\DeclareMathOperator*{\esssup}{ess\,sup}

\newlength\myindent
\begin{document}

\sectionhead{Contributed research article}
\volume{XX}
\volnumber{YY}
\year{20ZZ}
\month{AAAA}

\begin{article}
\title{\pkg{conformalInference.multi} and \pkg{conformalInference.fd}: Twin Packages for Conformal Prediction}
\author{by Paolo Vergottini, Matteo Fontana, Jacopo Diquigiovanni, Aldo Solari and Simone Vantini}

\maketitle

\abstract{
Building on top of a regression model, Conformal Prediction methods produce distribution-free prediction sets, requiring only \textit{i.i.d.} data. While R packages implementing such methods for the univariate response framework have been developed, this is not the case with multivariate and functional responses.
\CRANpkg{conformalInference.multi} and \CRANpkg{conformalInference.fd} address this void, by extending classical and more advanced conformal prediction methods like full conformal, split conformal, jackknife+ and multi split conformal to deal with the multivariate and functional case. 
The extreme flexibility of conformal prediction, fully embraced by the structure of the package, which does not require any specific regression model, enables users to pass in any regression function as input while using basic regression models as reference.
Finally, the issue of visualisation is addressed by providing embedded plotting functions to visualize prediction regions.
}

\section{Introduction}
\label{sec:int}

Given a regression method, which outputs the predicted value \( \hat{y}\), how do I define a prediction set offering coverage $\geq 1-\alpha$?
Recently, powerful methods that eliminate the need for specific distributions of the data have been developed, making the question more compelling
Predictive tools of this type belong to the family of nonparametric prediction methods, of which conformal prediction \citep[][for a review]{algolearn, review} methods are the most prominent members. Due to their wide-applicability and flexibility these methods are increasingly used in the practice. While there are packages available for other programming languages that perform conformal prediction for regression in univariate response and multivariate response cases (with some limitations), there is a serious limitation in terms of availability in R.
A notable exception, available on GitHub is represented by \pkg{conformalInference}, which provides conformal prediction for regression with univariate responses. Details may be found in \cite{tib}. Still, there is no R package that is able to deal with multivariate or functional response cases.  
This is where our packages \pkg{conformalInference.multi} and \pkg{conformalInference.fd} come into play.
Our packages move from \pkg{conformalInference} while keeping the same philosophy with regards to user freedom (this will be discussed in subsection \nameref{sec:multireg}) .
\\
Conformal Prediction theory may simply rely on exchangeable data, but this assumption is frequently discarded in favor of \textit{i.i.d.} data.
Therefore in the following discussion we will consider a set of values \( (x_{1},y_{1}), (x_{2},y_{2}), ... ,(x_{n},y_{n}) \sim i.i.d.\) , and define \( z_{i} = (x_{i},y_{i}) \; \forall i\). 
Through a generic regression method (e.g. ridge regression, random forest, neural network etc.) we can construct, with the set \(\{ z_{1}, ... ,z_{n}\}\) , a regression estimator function \(\hat{\mu}\). Given a new input value \(x_{n+1}\), the predicted response will be \(\hat{\mu} (x_{n+1})\).
In our discussion, three frameworks will appear:
\begin{enumerate}
\item Univariate response: $y \in \mathbb{R}$
\item Multivariate response: $y \in \mathbb{R}^{q}$
\item Multivariate functional response: $y \in \prod_{j=1}^{q} L^{\infty}({\Large \tau}_{j})$, where ${\Large \tau} _{j}$ is a closed and bounded subset of $\mathbb{R}^{d_{j}}, d_{j} \in \mathbb{N}_{>0}$. In particular $y_i$ is a function $i=1,...,q$
\end{enumerate}
To simplify notation we will generalize $y \in \mathcal{Y}$, keeping in mind all the due differences among the different contexts. 
\\
\textbf{Non-conformity score} is the cornerstone of the Conformal Prediction framework. 
Indeed, given a new value  \(z_{n+1}\), we can measure how unusual it is compared to all the other observations \( \{z_1, ... , z_n\}\)  through the medium of a non-conformity measure, a real-valued function  \(\mathcal{A} \) ,  returning  \(\mathcal{A}(\{z_1, ... , z_n\},z_{n+1})\) and assigning greater value to the most unusual points. In the regression framework  the test value is \( z_{n+1} = (x_{n+1}, y )\), where \(y\) is the candidate response at the input point \(x_{n+1}\). Let's consider the different scenarios:

\begin{itemize}
\item In the univariate case the classical non-conformity measure is the absolute value of residuals:
\begin{equation*}
\mathcal{A}(\{z_1, ... , z_n\},z_{n+1}) = |y - \hat{\mu} (x_{n+1})|
\end{equation*}

\item In the multivariate case there are a few options :
\begin{itemize}
\item \(A(\{z_1, ... , z_n\},z_{n+1}) = ||y - \hat{\mu} (x_{n+1})||_2\)  \hspace*{\fill} l2 norm
\item \(A(\{z_1, ... , z_n\},z_{n+1}) = (y - \hat{\mu} (x_{n+1}))S^{-1}(y_{n+1} - \hat{\mu} (x_{n+1}))^{T}\)\hspace*{\fill} Mahalanobis dist.
\item \(A(\{z_1, ... , z_n\},z_{n+1}) = max(y - \hat{\mu} (x_{n+1}))\) \hspace*{\fill} max 
\end{itemize}

\item In the functional case we decided to adopt the non-conformity score defined in \cite{fun2}:
	\begin{equation}
    \label{eq:ncmmax}
	\mathcal{A}(\{z_1, ... , z_n\} \}, z_{n+1})=  \sup_{j \in \{ 1, ..., q\} }  \left( \esssup_{t \in \mathcal{\Tau}_{j}}  \left|\frac{y^{j}(t) - [\hat{\mu}^{j}(x_{n+1})](t)}{s^{j}(t)}\right| \right)
	\end{equation}
\end{itemize}

In equation \eqref{eq:ncmmax} $q$ is the number of components of the multivariate functional response, $\Tau_{j}$ is the space over which the  j\textsuperscript{th} functional takes value,  $\hat{\mu}^{j}_{\mathcal{I}_{1}}$ is the j\textsuperscript{th} component of the regression estimator trained over $\mathcal{I}_{1}$, and $s^{j}(t)$ is a modulation function for dimension $j$ built with the observations in $\{z_1, ... , z_n\}$. This non-conformity metric descends from the supremum and incorporates a scaling factor along each component of the response. 
By taking the inverse of any non-conformity measure, we obtain a \textbf{Conformity Measure}, i.e. a real-valued function \(\mathcal{D} \) which computes how well an observation $z_{n+1}$ conforms with the rest of the dataset $\{z_1, ... , z_n\}$.
\\ 
In all the presented non-conformity scores we can exploit \textbf{modulation functions}, as in \eqref{eq:ncmmax}, whose role is to scale the conformity score by taking into account the variability along all directions, in such a way that prediction regions may adapt their width according to the
local variability of data.  If we restrict to a simple multivariate response, we can also consider modulation factors as  $s^{1}_{\mathcal{I}_{1}}, ... , s^{p}_{\mathcal{I}_{1}}$. Following the line of reasoning in \cite{fun2}, we considered three alternatives: \textit{identity}, where no scaling is applied, \textit{st-dev}, where we scale with the standard deviation every component of the response, and \textit{alpha-max}, where we ignore the most extreme observations and modulates data on the basis of the remaining observations. The latter is defined as:

\begin{equation}
s^{j}_{\mathcal{I}_{1}}(t) := \frac{max_{h\in \mathcal{H}_{1}} |y_{h,j}(t) - [\hat{\mu}_{\mathcal{I}_{1}}^{j} (x_{h,j})   ] (t)     |}    {\sum_{j=1}^{p}    \int_{\large{\tau}_{j}}    max_{h\in \mathcal{H}_{1}} |y_{h,j}(t) - [\hat{\mu}_{\mathcal{I}_{1}}^{j} (x_{h,j})   ] (t)     |dt                     }
\end{equation}
 where $\mathcal{H}_{1} := \{ h \in \mathcal{I}_{1} : sup_{j \in \{ 1, ... , q \}}( sup_{t\in \large{\tau}_{j} }  |y_{h,j}(t) - [\hat{\mu}_{\mathcal{I}_{1}}^{j} (x_{h,j})   ] (t)     |)  \leq \gamma\}$ and $\gamma $ is the $\lceil (|\mathcal{I}_{1}|+1)(1-\alpha) \rceil$ smallest value in the set  $ \{ sup_{j \in \{ 1, ... , q \}} (sup_{t\in \large{\tau}_{j} }  |y_{h,j}(t) - [\hat{\mu}_{\mathcal{I}_{1}}^{j} (x_{h,j})   ] (t)     |)  :  h \in \mathcal{I}_{1} \}$. 

\section{Conformal prediction theory}

We will start by reviewing the key aspects of Conformal Prediction theory for regression in the univariate response case, and then we will transition to multivariate and functional response frameworks.

\subsection{Full conformal}
\label{sec:fc}
The first prediction method we will discuss is \textbf{Full Conformal}. Essentially, for a given test observation $x_{n+1}$, it ranks how well a candidate point $z_{n+1} = (x_{n+1}, y)$ matches with all the rest of the data, selecting $y$ from a grid of candidates.

While theoretically this approach can be extended to multivariate response contexts (of dimension $q$), by considering a $q$-dimensional grid, in practice it scales poorly, having to test $|grid \; points| ^ {q} $ candidates.
Thus, the approach can only be applied when $q$ is fairly small. On top of that, a potential extension to the functional case suffers from a critical problem: one should consider as candidates all possible functions in $\prod_{j=1}^{q} L^{\infty}({\Large \tau}_{j})$, where ${\Large \tau} _{j}$ is a closed and bounded subset of $\mathbb{R}^{d_{j}}, d_{j} \in \mathbb{N}_{>0}$. \\
To compute all the non-conformity scores we need to construct an augmented regression estimator \( \hat{\mu}_{y}\) , trained over the augmented data set \( \{z_1, ... , z_n, z_{n+1} \}\),
 and choose the function $\mathcal{A}$.
With univariate data, we will consider the absolute value of the residuals, while with multivariate we can take the $l_2$ norm as an example. We define the residual's scores as:

\begin{equation}
\label{eqn: residuals}
R_{i}: = \mathcal{A}(\{z_1, ... , z_{i-1}, z_{i+1}  , ... ,z_n, ,z_{n+1}\},z_{i}) \; \;  \; R_{n+1}: = \mathcal{A}(\{z_1, ... ,z_n\},z_{n+1}) 
\end{equation}

Our next step is to rank \(R_{n+1}\) with respect to all the other residual scores \(R_{1}, ... ,R_{n}\) and calculate the fraction of values with higher non-conformity scores than the candidate point as

\begin{equation}
\label{eqn:deltafull}
\delta_{y} = \frac{| {j \in \{1, ... , n, n+1\} \; : R_{j} \geq R_{n+1}} |}{n+1}
\end{equation}

Lastly, we identify the full conformal prediction set as \(C_{full}(x_{n+1}) = \{ y\in \mathcal{Y} \; : \delta_{y} > \alpha \}\). 
As shown in \cite{distribfree}, the quantity \(1- \delta_{y}\) can serve as a valid p-value for testing the null hypothesis that \(H_{0} : Y_{n+1}=y\), and the following theorem holds:

\begin{equation}
\mathbb{P}(y_{n+1} \in C_{full}(x_{n+1})) \geq 1-\alpha
\end{equation}
Therefore, the method guarantees a coverage level of $1-\alpha$. The pseudo-code for the method can be found in Algorithm \ref{algofull}.


\begin{algorithm}[H]
\captionsetup{singlelinecheck = false, justification=justified}
\caption{Full Conformal Prediction}
\label{algofull}
\vspace{-3mm}
\hrulefill

\SetKwInput{KwInput}{Input}                
\SetKwInput{KwOutput}{Output}              
\DontPrintSemicolon
  
  \KwInput{level \(\alpha \in (0,1)\), and a regression algorithm $\hat{\mu}$ }
   \KwData{$ z_{i} = (x_{i},y_{i}) \; for \; i=1,...,n$ ,  test point $x_{n+1}$, and a set of candidate responses $\mathcal{Y}^{c} = \{ y_{1}, y_{2}, ... \}$}
  \KwOutput{Prediction intervals at point $x_{n+1}$}
         \ForEach{$y \in \mathcal{Y}^{c}$}
    {
       $\hat{\mu}_{y} = \hat{\mu}(z_{1}, ... , z_{n}, (x,y)) $ \\
       Compute nonconformity scores: \\
       $R_{i} = |y_{i} - \hat{\mu}_{y}(x_{i}) |\; \; and \; \; R_{n+1} = |y - \hat{\mu}_{y}(x) | $  \; (Univariate)\\
       $R_{i} = ||y_{i} - \hat{\mu}_{y}(x_{i}) ||_{2}\; \; and \; \; R_{n+1} = ||y - \hat{\mu}_{y}(x_{n+1}) ||_{2} $ \; (Multivariate)\\

       $\delta_{i} = \frac{1+\sum_{k=1}^{n} \mathbbm{1}(R_{i}) \leq R_{n+1}}{n+1}$
    }
   
\Return $C_{full}(x_{n+1}) = \{ y \in \mathcal{Y}^{c} : \delta_{y}\geq \alpha  \}$

\end{algorithm}

\subsection{Split conformal}
\label{sec:sc}
As already stated, Full Conformal prediction has computational issues that prevent it to be used in cases for which recomputing the regression model is costly, or when, due to the dimensionality of the data, the number of points of the grid to be explored becomes simply too big. For this reason \textbf{Split Conformal Prediction}, also mentioned as Inductive (with respect to the full or transductive conformal prediction) was suggested to overcome the high computational load of full conformal. 

In mathematical terms, the observations $y_{1}, ..., y_{n}$ are split into a training set $\mathcal{I}_{1}$ and a validation set  $\mathcal{I}_{2}$, respectively of cardinality $m$ and $l$. 
The first set is used to train, just once, the regression model, while the second set is used to compute, for every $ i \in \mathcal{I}_{2}$, the distance of each $y_{i} $ from the fitted value $\hat{\mu}_{\mathcal{I}_{1}}(x_{i})$ through a non-conformity measure. 
Indeed, consider 

\begin{equation}
\label{eqn: residuals}
R_{i}: = \mathcal{A}(\{z_k : k \in \mathcal{I}_1\},z_{i}) \; \; \; i \in \mathcal{I}_2\; \; \;  \;  \; R_{n+1}: = \mathcal{A}(\{z_k : k \in \mathcal{I}_1\},z_{n+1}) 
\end{equation}

and use the new definitions of residuals' scores in \eqref{eqn: residuals} to obtain $\delta_i$ as in \eqref{eqn:deltafull}. \\
Hence, $C_{split}(x_{n+1}) = \{ y\in \mathcal{Y} \; : \delta_{y} > \alpha \}$. As shown by \cite{algolearn}, the following inequalities state the theoretical coverage of the split conformal method:

\begin{equation}
1-\alpha \leq \mathbb{P}(y_{n+1} \in C_{split}(x_{n+1}) ) < 1-\alpha + \frac{1}{l+1}
\end{equation}
Thus, split conformal prediction sets are valid, but there is no guarantee of exactness, i.e. having precise coverage of $1-\alpha$. 
A slight modification can provide exactness: if we introduce a randomization element $\tau_{n+1} \sim \mathcal{U}_{[0,1]}$, we can express the smoothed split conformal predictions set $C_{split, \tau_{n+1}}(x_{n+1}) := \{ y \in \mathcal{Y} : \delta_{y, \tau_{n+1}} > \alpha   \}$  , where

\begin{equation}
\delta_{y, \tau_{n+1}} = \frac{|\{j \in \mathcal{I}_{2} : R_{j} > R_{n+1}\}| + \tau_{n+1} |\{ j\in \mathcal{I}_{2} \cup \{ n+1\} : R_{j} = R_{n+1}  \}|}{l+1}
\end{equation}

The proof showing that the smoothed version leads to exact prediction sets for any value of $\alpha$ and $l$ is in \cite{fun1}.
\\ 
According to another formulation of split conformal one might define $k = \lceil (1-\alpha) (l+1)\rceil$ and select the $k^{th}$ smallest non-conformity score in $\{ R_i \; : i \in \mathcal{I}_2\}$ as $d$.
Then the alternative definition of the split conformal prediction set is  

\begin{equation}
    C_{split} (x_{n+1}) =  [\hat{\mu}_{\mathcal{I}_{1}} (x_{n+1}) - d , \hat{\mu}_{\mathcal{I}_{1}} (x_{n+1}) + d ]
\end{equation}

Likewise, for the smoothed version one may define $k_{\tau_{n+1}} = \lceil l + \tau_{n+1}-(l+1)\alpha\rceil$ and select the $k_{\tau_{n+1}}-th$ smallest non-conformity score in $\{ R_i \; : i \in \mathcal{I}_2\}$ as $d_{\tau_{n+1}}$.
Thus, another way of defining the split conformal prediction set would be:

\begin{equation}
    C_{split, \tau_{n+1}} (x_{n+1}) =  [\hat{\mu}_{\mathcal{I}_{1}} (x_{n+1}) - d_{\tau_{n+1}} , \hat{\mu}_{\mathcal{I}_{1}} (x_{n+1}) + d_{\tau_{n+1}} ]
\end{equation}

Observe that if $\tau_{n+1}=1$, we revert back to the classical split conformal version. The pseudo-code is shown in Algorithm \ref{alg:algosplit}.
\\
As long as a suitable non-conformity measure is employed, the split conformal method can be applied successfully in the multivariate response and multivariate functional response frameworks as well.

\begin{algorithm}[H]
\captionsetup{singlelinecheck = false, justification=justified}
\caption{Split Conformal Prediction}
\label{alg:algosplit}

\vspace{-3mm}
\hrulefill

\SetKwInput{KwInput}{Input}                
\SetKwInput{KwOutput}{Output}              
\DontPrintSemicolon
  
  \KwInput{split proportion $\rho$, level \(\alpha \in (0,1)\), and a regression algorithm $\hat{\mu}$ }
   \KwData{$ z_{i} = (x_{i},y_{i}) \; for \; i=1,...,n$ ,  and a test point $x_{n+1}$  }
  \KwOutput{Prediction sets at each point in $\mathcal{X_{0}}$}
  \: Randomly split $\{ 1, ... ,n\}$ into $\mathcal{I}_{1}$ and $\mathcal{I}_{2}$ so that $|\mathcal{I}_{1}| = \rho n$ and $|\mathcal{I}_2| = \rho(1-n) = : l$ \\
  Build the regression function $\hat{\mu}_{\mathcal{I}_{1}} = \hat{\mu}(\{z_{k} \; : \; k\in \mathcal{I}_{1}\})$ \\
   Compute nonconformity scores for $i \in \mathcal{I}_{2}$ : \\
   \hspace*{3mm } $R_{i} = |y_{i} - \hat{\mu}_{\mathcal{I}_{1}}(x_{i})|$ \; (Univariate)\\
      \hspace*{3mm } $R_{i} = ||y_{i} - \hat{\mu}_{\mathcal{I}_{1}}(x_{i})||_{2}$ \; (Multivariate)\\
     \hspace*{3mm } $R_{i} = \sup_{j \in \{ 1, ..., q\} } \left ( \sup_{t \in \mathcal{\Tau_{j}}}  \left|\frac{y_{i,j}(t) - [\hat{\mu}^{j}_{\mathcal{I}_{1}}(x_{i,j})](t)}{s^{j}_{\mathcal{I}_{1}}(t)}\right| \right )$ \; (Functional) \\
     Now choose between smoothed or classical version:      \vspace*{1mm}
\\
       \hspace*{3mm }  $k = \lceil (l+1)(1-\alpha) \rceil$  (Classical)\\
      \hspace*{3mm } Sample $\tau_{n+1} \sim \mathcal{U}_{[0,1]}$ and   $k = \lceil l + \tau_{n+1}-(l+1)\alpha\rceil$ (Smoothed) \vspace*{1mm}\\
  d = k-th smallest value in $\{  R_{k} \; : \; k \in \mathcal{I}_{2} \}$ 

\vspace*{2mm}
\Return  $C_{split} (x_{n+1}) =  [\hat{\mu}_{\mathcal{I}_{1}} (x_{n+1}) - d , \hat{\mu}_{\mathcal{I}_{1}} (x_{n+1}) + d ] $

\end{algorithm}

\subsection{Jackknife+}
\label{sec:jack}

The \textbf{Jackknife+} method was introduced in \cite{jackplus} as a compromise between the full conformal and split conformal complexity. This method relies as well on the assumption of \textit{i.i.d.} data.

As this is a slight variation of the well-known Jackknife method, we should first review the classic jackknife prediction interval.
Indeed, assuming $(x_1, y_1),..,(x_n, y_n) \;\; i.i.d$ :

\begin{equation}
\label{eq:jack}
    C_{jack}=[q_{\alpha}\{\hat{\mu}(x_{n+1}) - R_{i}^{LOO}\}, \;
    q_{1-\alpha}\{\hat{\mu}(x_{n+1}) + R_{i}^{LOO}\}]
\end{equation}

where  $R_{i} = |y_{i}-\hat{\mu}_{-i}(x_{i})| \; \forall i=1, ..., n$ are the absolute leave-one-out (or LOO) residuals and $\hat{\mu}$ is a regression model trained over the whole dataset $\{z_1,...,z_n \}$.
The LOO residuals can be computed by fitting a complete regression model and $n$ leave-one-out models, removing one observation at a time for training. It provides symmetric intervals, since they are all centered around $\hat{\mu}(x_{n+1})$, and it tackles the issue of overfitting by using the quantile of the absolute residuals from LOO. Although fitting $n+1$ models seems extremely challenging, some models have shortcuts for obtaining LOO residuals (for example, the linear model can use the Sherman-Morrison formula).

In terms of coverage, jackknife is lacking sufficient theoretical guarantees. Indeed one can construct an extreme case where \eqref{eq:jack} leads to empty coverage.
As an alternative to jackknife, jackknife+ is very appealing: it provides a guarantee of $1-2\alpha\%$ coverage for a slightly higher computational demand.
The complete proof is discussed in \cite{jackplus}.
\\
Let us now present the jackknife+ interval formulation:

\begin{equation}
\label{eq:jackplus}
    C_{jack+}=[q_{\alpha}\{\hat{\mu}_{-i}(x_{n+1}) - R_{i}^{LOO}\}, \;
    q_{1-\alpha}\{\hat{\mu}_{-i}(x_{n+1}) + R_{i}^{LOO}\}] 
\end{equation}

Unlike \eqref{eq:jack}, the predictions are not centered on $\hat{\mu}(x_{n+1})$ but instead are shifted by $\hat{\mu}_{-i}(x_{n+1})$. This is most noticeable when the regression model is highly sensitive to training data.
As previously mentioned we can show that:

\begin{equation}
    \mathbb{P}(y_{n+1}\in C_{jack+}(x_{n+1}) ) \geq 1-2\alpha
\end{equation}

Since jackknife+ requires a univariate quantile, it is not straightforward to translate that concept to a multivariate or functional model. A possible solution revolves around the concept of conformity measure $\mathcal{D}$, introduced in section \nameref{sec:int}.
The conformity score computed by $\mathcal{D}$ indicates how well a data point fits with the rest of the data. A point that does not match the rest will present a low conformity score, while one that does will present a high conformity score.
As a measure of conformity for both multivariate and functional responses, we chose:

\begin{equation}
\begin{split}
  \mathcal{D}_{max}(x,y) &= \left( \sup_{j \in \{ 1, ..., q\} }  \left|\frac{y_{j}- [\hat{\mu}^{j}(x)]}{s^{j}}\right| \right)^{-1} (multivariate) \\
  &= \left[ \sup_{j \in \{ 1, ..., q\} } \left ( \esssup_{t \in \mathcal{\Tau}_{j}}  \left|\frac{y_{j}(t) - [\hat{\mu}^{j}(x_{j})](t)}{s^{j}(t)}\right| \right) \right]^{-1}  (functional) 
\end{split}
\label{eq:depthmax}
\end{equation}

%

For the sake of simplicity, we will refer to $\mathcal{D}_{max}(x,y)$, while keeping in mind the two different formulations.
In order to replicate the concept of quantile, we ordered the data points according to their conformity measures.
This allows us to select as a multivariate and functional quantile ($q_{\alpha}^{\mathcal{D}}$), the level set induced by the conformity measure. In particular consider $u_1, .., u_n \in \mathcal{U}$, then:

\vspace*{-3mm}
\begin{equation}
\label{eq:extquant}
q_{\alpha}^{\mathcal{D}}(u_1, .., u_n): = \{u\in \mathcal{U} : \mathcal{D}_{max}(u) \geq q_{\alpha}\{\mathcal{D}_{max}(u_1), ...,\mathcal{D}_{max}(u_n) \}\}
\end{equation}

Note $q_{\alpha}\{\mathcal{D}_{max}(u_1), ...,\mathcal{D}_{max}(u_n) \}$
refers to the classical quantile.
Moreover, this conformity measure also mirrors the behaviour of depth: in addition to the rating from unusual to more "common" data, it also provides an outward-centered ranking, with low scores assigned to extreme data.
In any case, the ranking induced by \eqref{eq:extquant} is entirely different from the smallest-to-largest values ranking in the univariate case. Consequently, the notion of extended quantile should take into account this interpretation change. $q_{\alpha}^{\mathcal{D}}$ is not always a "small" value (in fact, the concept of being small is meaningless in this framework), but rather an extreme or less conformal value.
The converse reasoning can be applied to
$q_{1-\alpha}^{\mathcal{D}}$.
\\
As a result, in the multivariate context we could simply extend \eqref{eq:jackplus}  as :

\begin{equation}
C_{jack+}= \{ y \in
\mathbb{R}^q : y  \in [q^{\mathcal{D}}_{\alpha}(\{\hat{\mu}_{-i}(x_{n+1}) \pm R_{i}^{LOO} : i=1,...,n\})] \}
\end{equation}

\hspace{-4.5 mm}
While in the functional context:

\begin{equation}
C_{jack+}= \{ y \in \prod_{j=1}^{q} L^{\infty}({\Large \tau}_{j}) : y(t)  \in [q_{\alpha}^{\mathcal{D}}(\{\hat{\mu}_{-i}(x_{n+1}) \pm R_{i}^{LOO} : i=1,...,n\})(t) ] \; \forall t \in \prod_{j=1}^{q} \tau_{j}  \}
\end{equation}

As shown in Algorithm \ref{alg:jackp}, after obtaining the level sets with $q_{\alpha}^{\mathcal{D}}$, we compute the axis-aligned minimum bounding box (or AABB), effectively projecting the prediction region over the axis. The method yields a prediction interval for each component of the response, simplifying the interpretation of the results.

\begin{algorithm}[H]
\captionsetup{singlelinecheck = false, justification=justified}
\caption{Jackknife+ Prediction}
\label{alg:jackp}

\vspace{-3mm}
\hrulefill

\SetKwInput{KwInput}{Input}                
\SetKwInput{KwOutput}{Output}              
\DontPrintSemicolon
  
  \KwInput{level \(\alpha \in (0,1)\), and a regression algorithm $\hat{\mu}$ }
   \KwData{$ z_{i} = (x_{i},y_{i}) \; for \; i=1,...,n$ , and a test point $x_{n+1}$  }
  \KwOutput{Prediction set at point $x_{n+1}$}
  
  \For{i=1, ... , n}{
  Train regression over \(\{z_{1}, ... , z_{i-1}, z_{i+1}, ..., z_{n}\}\) obtaining $\hat{\mu}_{-i}$ \\
  Compute LOO residual $R_{i}^{LOO} = |y_{i}-\hat{\mu}_{-i} (x_{i})|$
  }
  $R^{-}:=\{\hat{\mu}_{-i}(x_{n+1}) - R_{i}^{LOO}\ \; i=1,...,n \}$ \\ \hspace*{-2.5 mm}
  $R^{+}:=\{\hat{\mu}_{-i}(x_{n+1}) + R_{i}^{LOO}\ \; i=1,...,n \}$ \\ \hspace*{-2.5 mm}
  $R^{\pm}:=\{\hat{\mu}_{-i}(x_{n+1}) \pm R_{i}^{LOO}\ \; i=1,...,n \}$ \\
  
  \(C_{jack+}(x_{n+1})=[q_{\alpha}(R^{-}), q_{1-\alpha}(R^{+}) ]\) (Univariate)
 
    $C_{jack+}(x_{n+1})= Bounding Box (q^{\mathcal{D}}_{\alpha}(R^{\pm}))$ (Multivariate - Functional)
  \\
  \hspace*{-2.5mm}
  \KwRet $C_{jack+}(x_{n+1})$

\end{algorithm}

\subsection{Multi split conformal}
\label{sec:msc}
Among the drawbacks of split conformal is the inherent randomness in the splitting procedure, since each split produces a valid prediction set. To overcome this limitation, the \textbf{Multi Split Conformal} method was developed. As described in \cite{msplit}, the split conformal method is run multiple times ($B$) and then the prediction intervals are combined. To determine the minimum number of intervals within which a point must be contained to be included in our final prediction set, we use $\tau$. Formally, given the split conformal intervals $C^{ [1]}, ..., C^{ [B]}$, we can define $\Pi ^{y} =  \frac{1}{B} \sum_{b=1}^{B} \mathbb{1} \{ y\in C^{[b]} \} \; \forall y \in \mathbb{R}$. Then the multi split conformal prediction interval is:

\begin{equation}
C_{msplit} (x_{0}) = \{ y \in \mathbb{R} \; : \; \Pi^{y} > \tau \}
\end{equation}

It is important to note that the split conformal intervals $C^{ [1]}, ..., C^{ [B]}$ are obtained by setting the miscoverage probability to $\alpha(1-\tau+\lambda/B)$, where $\lambda$ is a smoothing parameter (a positive integer). As proven in \cite{msplit}, the multi split prediction interval has coverage at least $1-\alpha$.
\\
The extension to the multivariate and functional case may be inspired by the one discussed in the previous subsection. Indeed, one runs the split conformal methods multiple times and joins their lower bounds and upper bounds into a single set. 
The next step employs the notion of quantile defined in \eqref{eq:extquant}.
By considering only points with significant conformity scores, we can build a level set induced by the ranking provided by $\mathcal{D}^{max}$. This time around, we will choose the level of the quantile based on $2\tau B$, i.e. twice the chosen amount of areas or bands a point must be contained in to be part of the final prediction set.
For the final prediction set, we consider the axis-aligned bounding box of the produced output, just as we did with the jackknife+ extension.
An efficient way to compute the multi split prediction interval, taken from \cite{gupta}, is shown in Algorithm \ref{alg:msplit}.

\begin{algorithm}[H]
  \SetAlgoLined
\captionsetup{singlelinecheck = false, justification=justified}
 \caption{Multi Split Conformal Prediction}

  \label{alg:msplit}
  
  \vspace{-3mm}
\hrulefill
\SetKwInput{KwInput}{Input}                
\SetKwInput{KwOutput}{Output}              
\DontPrintSemicolon
  
  \KwInput{split proportion vector $prop$, level \(\alpha \in (0,1)\), and a regression algorithm $\hat{\mu}$, number of replications B, smoothing parameter $\lambda$, joining parameter $\tau$ }
  \KwOutput{$ z_{i} = (x_{i},y_{i}) \; for \; i=1,...,n$ ,  and a test point $x_{n+1}$}
  \KwData{Testing set $x$}
  \For{b = 1, ... , B}{
  		$(C^{ [b]},lo^{[b]},up^{[b]})$=Split($\{z_{1}, ... ,z_{n} \} $, $x_{n+1}$, $prop[b]$, $\alpha(1-\tau+\lambda/B)$, $\hat{\mu}$ )
  }
  \If{univariate case}{
   $\Pi ^{y} =  \frac{1}{B} \sum_{b=1}^{B} \mathbb{1} \{ y\in C^{[b]} \} \; \forall y \in \mathbb{R}$ \\ \hspace*{-2.5mm}
$C_{msplit} (x_{n+1}) = \{ y \in \mathbb{R} \; : \; \Pi^{y} > \tau    \}$
  
  }
  \If{multivariate-functional case}{
  $L=\{ lo^{[b]}, up^{[b]} \; b=1,...,B\}$ \\
  \hspace*{-2.5mm} $L_q = q^{\mathcal{D}}_{2\tau B}(L)$ \\
  \hspace*{-2.5mm}  $C_{msplit}(x_{n+1})= Bounding Box (L_q)$
  
  }
  \KwRet $C_{msplit}(x_{n+1})$
  
  \vspace{2mm}

  \SetKwFunction{FSplit}{Split}
 
  \SetKwProg{Fn}{Function}{:}{}
  \Fn{\FSplit{$\{z_{1}, ... ,z_{n} \} $, $x_{0}$, $\rho$, $\alpha$, $\hat{\mu}$ }}{
   \;     Randomly split $\{ 1, ... ,n\}$ into $\mathcal{I}_{1}$ and $\mathcal{I}_{2}$ so that $|\mathcal{I}_{1}| = \rho n$ \\
  Build the regression function $\hat{\mu}_{\mathcal{I}_{1}} = \hat{\mu}(\{z_{k} \; : \; k\in \mathcal{I}{1}\})$ \\
  $R_{i} = |y_{i} - \hat{\mu}_{\mathcal{I}_{1}}(x_{i})|$ \\
  $k = \lceil n(1-\rho)(1-\alpha) \rceil$ \\
  d = k-th smallest value in $\{  R_{k} \; : \; k \in \mathcal{I}_{2} \}$

\KwRet $C_{n,1-\alpha}^{split} (x_{n+1}) =  [\hat{\mu}_{\mathcal{I}_{1}} (x_{n+1}) - d , \hat{\mu}_{\mathcal{I}_{1}} (x_{n+1}) + d ] $, \\ \hspace*{8mm} $lo = \hat{\mu}_{\mathcal{I}_{1}} (x_{n+1}) - d$, $up = \hat{\mu}_{\mathcal{I}_{1}} (x_{n+1}) + d$  
  }
  \;
 
\end{algorithm}

\section{conformalInference.multi : structure and functionality}
\label{sec:multiint}
The \pkg{conformalInference.multi} package performs conformal prediction for regression when the response variable $y$ is multivariate. 
Table \ref{tablemulti} contains a list of all available functions. Additionally, there are a few helper functions implemented in the package: for instance, the \texttt{check} functions verify the type correctness of input variables (if not, a warning is displayed to the user and the code execution is interrupted), \texttt{interval.build}  joins multiple regions into a single one and is vital to the implementation of the multi split method, and \texttt{computing\_s\_regression} provides modulation values for the residuals in the conformal prediction methods. \\
One can mainly identify two types of methods: regression methods and prediction functions. 

\begin{table}[h!]
\centering

 \begin{tabular}{|c c |} 
 \hline 
 Function & Description \\
 \hline\hline
\texttt{ conformal.multidim.full} & Computes Full Conformal prediction regions \\
\texttt{conformal.multidim.jackplus} & Computes Jackknife+ prediction regions \\
{ \texttt{conformal.multidim.split}} & Computes Split Conformal prediction regions \\
{ \texttt{conformal.multidim.msplit}} & Computes Multi Split Conformal prediction regions \\
{ \texttt{elastic.funs}} & Build elastic net regression \\
{\texttt{lasso.funs}} &Build lasso regression \\
\texttt{lm\_multi} & Build linear regression \\
\texttt{mean\_multi} & Build regression functions with mean  \\
\texttt{plot\_multidim} & Plot the output of prediction methods \\
{\texttt{ridge.funs}} & Build elastic net regression \\
 \hline
 \end{tabular}
 \caption{List of the main functions in \pkg{conformalInference.multi}. Helpers functions are excluded. } \label{tablemulti}
\end{table}

\subsection{Regression methods}
\label{sec:multireg}

The philosophy behind this package is based on one major assumption: the regression methods should not be included in the prediction functions themselves, as the user may require fitting a very specific regression model. Because of this, we let the final user create his or her own regression method by mimicking the proposed regression algorithms and passing it as input to the prediction algorithms. Nevertheless, for users with less experience or less demanding needs, some regression methods have already been introduced.
\\
To keep the problem simple, we assumed each dimension of the response was independent, so each component was fitted with a separate model. Although this may seem restricting, native models can be built for multivariate multiple regression.
Three different methods are available: the \textit{mean model}, where the response is modeled using the sample mean of the observations, the \textit{linear model} and the \textit{elastic net model}, which applies the \CRANpkg{glmnet} package and also contains lasso and ridge regression subcases. Two elements will be returned by these functions: \texttt{train.fun} and \texttt{pred.fun}. The first, given as input the matrices $x$ and $y$ (avoid data frames, as depending packages may not accept them), provides the training function, while the second, taking as input the test values and the output of \texttt{train.fun}, returns the predictions at the test values $x_0$.

\subsection{Prediction methods}
\label{sec:multipre}

All the prediction functions take as input a matrix \texttt{x} of dimension $n$ x $p$, a matrix of responses \texttt{y} of dimension $n$ x $q$, a test matrix \texttt{x0} with dimension $n_0$ x $p$ and the level of the prediction $\alpha \in (0,1)$.
With the string \texttt{score} we can select one of three non-conformity scores: \textit{l2, mahalanobis, max}. Moreover, the scores can be scaled either using the \textit{mad functions} or with the parameter  \texttt{s.type}. The \textit{mad functions}, i.e. \texttt{mad.train.fun} and \texttt{mad.predict.fun}, are user defined functions (with the same structure as the regression methods) that scale the residuals, whereas the parameter \texttt{s.type} determines a modulation function between \textit{identity}, no residual scaling, \textit{st-dev}, dividing the residuals by the standard deviation along various dimensions and \textit{alpha-max}, which produces modulations similar to those of \textit{st-dev}, but ignores  the more extreme observations for the local scaling.
In addition, a logical \texttt{verbose} allows to print intermediate processes.
\CRANpkg{future.apply}, which leverages \CRANpkg{future} to parallelize computations, was used to speed up computations in all cases, except for the split conformal prediction method.
\\ \\
The first implemented method is full conformal. The algorithm requires a grid of candidate response for each test value in \texttt{x0}. The grid points in dimension $k$ are \texttt{num.grid.pts.dim} points equally spaced in the interval
\([-grid.factor * max|y(k)| , grid.factor * max|y(k)] ]\), where $y(k)$ is the value of the response in the k\textsuperscript{th} dimension. For this range of trial values (assuming $grid.factor \geq 1$) the cost in coverage is at most $\frac{1}{n+1}$, as shown in \cite{trim}. 
Additionally, in full conformal the \textit{alpha-max} option for \texttt{s.type} is not included. The function returns the predicted values and a list of candidate responses at each element in $x_0$.

\begin{example}
conformal.multidim.full = function(x, y, x0, train.fun, predict.fun,alpha = 0.1,
				mad.train.fun = NULL,mad.predict.fun = NULL,
                                score='l2', s.type = "st-dev",
                                num.grid.pts.dim=100, grid.factor=1.25,
                                verbose=FALSE)
\end{example}

Alternatively, the split conformal method divides the dataset into a training set and a validation set, and this division is controlled by the argument \texttt{split}, a vector of indices for the training set, or by the argument \texttt{seed}, which initializes a pseudo random number generator for splitting. A randomized version of the algorithm can be selected using \texttt{randomized}, and the sampled value of $\tau$ can be modified by \texttt{seed.rand}. By adjusting \texttt{rho}, the user can also tune the ratio between train and validation. The split function outputs the predicted values and prediction intervals for every component at \texttt{x0}.

\begin{example}
conformal.multidim.split = function(x,y, x0, train.fun, predict.fun, alpha=0.1,
                                 split=NULL, seed=FALSE, randomized=FALSE,seed.rand=FALSE,
                                 verbose=FALSE, rho=0.5,
                                 score="l2",s.type="st-dev", mad.train.fun = NULL,
                                 mad.predict.fun = NULL)
                                 
\end{example}

In the jackknife+ prediction function, the LOO residuals and $n$ models are computed to obtain fitted values at each point in \texttt{x0}. Accordingly, an extended quantile is obtained by using the conformity measure in \eqref{eq:extquant} and used to build the lower and upper prediction intervals for every component, which will be returned as output.

\begin{example}
conformal.multidim.jackplus = function(x,y,x0, train.fun, predict.fun, alpha=0.1)
\end{example}

Multi split conformal prediction is the last prediction function, where \texttt{B} is the number of times the split conformal function is run. As explained in subsection \nameref{sec:msc}, \texttt{lambda} controls the smoothing for roughness penalization, while the argument \texttt{tau} controls the miscoverage level for the split conformal functions as well as how many points are considered by the extended quantile in \eqref{eq:extquant}. 
\texttt{B} is set to 100 by default, while \texttt{tau} is set to 0.1. The output is identical to \texttt{conformal.multidim.jackplus}.

\begin{example}
conformal.multidim.msplit = function(x,y, x0, train.fun, predict.fun, alpha=0.1,
                                      split=NULL, seed=F, randomized=F,seed.rand=F,
                                      verbose=F, rho=NULL,score = "max",
                                      s.type = "st-dev",B=100,lambda=0,
                                      tau = 0.1 ,mad.train.fun = NULL,
                                      mad.predict.fun = NULL)
\end{example}

The results of a prediction method are then plotted using the function \texttt{plot\_multidim} that takes as input \texttt{out}, the output of the prediction method, and a logical \texttt{same.scale} that forces the y-axes to utilize the same scale for each component. The function depends on the packages \CRANpkg{ggplot2}, \CRANpkg{gridExtra} and \CRANpkg{hrbrthemes}. Specifically, if I pass the output of full conformal, it plots a heatmap for each test point in \texttt{x0}. The color intensifies whenever the estimated p-value increases.
 Conversely, if the input is the result of the split, multi split or jackknife+, then it plots confidence intervals for each observation by pairing a component for the x value and a component for the y value.

\begin{example}
plot_multidim=function(out,same.scale=FALSE)
\end{example}

\section{conformalInference.fd : structure and functionality}
\label{sec:fdint}

\pkg{conformalInference.fd} provides conformal prediction when the response variable $y$ is a multivariate functional datum.
In the same manner as \pkg{conformalInference.multi}, a table of all the available functions is shown in Table \ref{tablefda}. A number of helper functions are not present in the table, including the \texttt{check} functions, which verify the type correctness of input variables, \texttt{convert2data} and \texttt{fun2data}, which convert functional inputs (of type fda, fData or mfData) to a list of points, \texttt{table2list}, which trasforms matrices to lists. \\
Moreover, this package also shares the structure with its multivariate counterpart: it is organized into regression methods and prediction functions.

\begin{table}[h!]
\centering

 \begin{tabular}{|c c |} 
 \hline 
 Function & Description \\
 \hline\hline
 \texttt{concurrent} & Build concurrent regression model \\
 { \texttt{conformal.fun.jackplus}} & Computes Jackknife+ prediction sets \\
{ \texttt{conformal.fun.split}} & Computes Split Conformal prediction sets \\
{ \texttt{conformal.fun.msplit}} & Computes Multi Split Conformal prediction sets \\
\texttt{mean\_lists} & Build regression method with mean \\
\texttt{plot\_fun} & Plot the output prediction methods \\
 \hline
 \end{tabular}
 \caption{List of the main functions in \pkg{conformalInference.fd}. Helpers functions are excluded. }
 \label{tablefda}
\end{table}

\subsection{Regression methods}
\label{sec:fdreg}
Because users may require novel regression models in the future, this package allows them to build their own methods and pass them as arguments to the prediction function. For simpler analysis we just implemented two elementary models for functional regression: \textit{mean model}, where y is modeled as the mean of the of the observed responses, and a \textit{concurrent regression model}, with the form:

\[y_{k}(s) = \beta_{0}(s) + \sum_{i=1}^{p}\beta_{i}(s)x_{i}(s)+\epsilon(s) \; \; k=1, ..., q\]

For this latter model, the evaluation grid for function $x \in \prod_{j=1}^{p} L^{\infty}({\Large \tau}_{j})$ and for $y \in \prod_{j=1}^{q} L^{\infty}({\Large \tau}_{j})$ must be the same.
These regression functions will return a list containing two functions: \texttt{train.fun}, which requires as input the lists \texttt{x},\texttt{t}, and \texttt{y}, and \texttt{pred.fun}, which takes as input the test values \texttt{x0} and the output of \texttt{train.fun}.

\subsection{Prediction methods}
\label{sec:fdpre}

When dealing with functional data, the input structure tends to be quite complex. To simplify the input, we developed \texttt{convert2data}, a function that converts functional data types into lists of punctual evaluations. 
The prediction functions make use of the following inputs: \texttt{x}, the regressors,  \texttt{t\_x}, the grid over which $x$ is evaluated,\texttt{y}, the responses, \texttt{x0}, the test values, \texttt{t\_y}, the grid over which \texttt{y} is evaluated.
\texttt{t\_x} and \texttt{t\_y} are lists (length $p$ and $q$ respectively) of vectors.
As previously mentioned, it is possible to define \texttt{x},\texttt{y} and \texttt{x0} as functional data of types $fD$, $fData$ or $mfData$, or to specify them as lists of vectors: the external list describes the observation, the inner one defines the components, while the final vector contains the punctual evaluation of an observation on a specific component (the number of evaluated points may differ). To be more clear:
\begin{itemize}
\item $x \rightarrow$ list (length \(n\)) of lists (length \(p\)) of vectors (arbitrary length)
\item $y \rightarrow$ list (length \(n\)) of lists (length \(q\)) of vectors (arbitrary length)
\item $x_0 \rightarrow$ list (length \(n_0\)) of lists (length \(p\)) of vectors (arbitrary length)
\end{itemize}

Additionally, the non-conformity score is set to the one presented in \eqref{eq:ncmmax} and the miscoverage level of the prediction is $\alpha$ $\in (0,1)$.
The division into train and validation is governed by the arguments \texttt{splits}, the vector of indices in training set, \texttt{seed}, i.e. the seed for the random engine, and $\rho \in (0,1)$, which represents the split proportion among training and validation sets.
One can select the randomized version of the algorithm with \texttt{randomize}, while the smoothness value is sampled from an $\mathcal{U}_{[0,1]}$ using as seed \texttt{seed.rand}.
 The user may select a \texttt{verbose} version which prints intermediate processes as they occur. With \texttt{s.type}, the user has a choice of three modulation methods: \textit{identity}, \textit{st-dev}, and \textit{alpha-max}, that are described in section \nameref{sec:int}.
All the methods return as output a list containing the evaluation grid for the response \texttt{t}, the lower and upper bounds \texttt{lo} and \texttt{up}. In particular the bounds are lists (of length $n_0$) of lists (of length $p$) of vectors. 
We again used \pkg{future.apply} for minimizing computation times, by parallelising the code.
\\  \\
First, we implement the split conformal function. This function generates, as well as the classical outputs, also the predicted values at \texttt{x0}. When no value is passed in for \texttt{x0}, then the function will use the mean as regression method and output the prediction bands at the validation set.
Following Algorithm \ref{alg:algosplit}:

\begin{example}
conformal.fun.split = function(x, t_x, y,t_y, x0, train.fun, predict.fun, alpha=0.1,
                             split=NULL, seed=FALSE, randomized=FALSE,seed.rand=FALSE,
                                verbose=FALSE, rho=0.5,s.type="st-dev") 
                                 
\end{example}

The second implemented method is jackknife+. Here, we compute the LOO residuals and $n$ models, removing one observation at a time.
As described in Algorithm \ref{alg:jackp}, we use the extended quantile defined in \eqref{eq:extquant} to compute the level set of the conformity score. The level set contains all the functional data points needed to build a prediction set. 
The last step involves computing an axis-aligned bounding box (AABB)  to determine a predicted set using the functional bands.

\begin{example}
conformal.fun.jackplus = function(x,t_x, y,t_y, x0, train.fun, predict.fun, alpha=0.1)
\end{example}

Alternatively, the multi split conformal function runs \texttt{B} times \texttt{conformal.fun.split} and joins the B prediction regions into one with the quantile formula in \eqref{eq:extquant}, selecting the $2\tau B$ most conformal bands.
Lastly, as with jackknife+, the AABB is returned.
We recommend you tune the value of \texttt{tau} for the particular problem at hand. Our default value is 0.5. 
\texttt{lambda} controls smoothing in the roughness penalization loss. Further details can be found in subsection \nameref{sec:msc}.

\begin{example}
conformal.fun.msplit = function(x,t_x, y,t_y, x0, train.fun, predict.fun, alpha=0.1,
                                split=NULL, seed=FALSE, randomized=FALSE,seed.rand=FALSE,
                                verbose=FALSE, rho=NULL,s.type="st-dev",B=50,lambda=0,
                                tau = 0.5) 
\end{example}

Our last effort was to design a versatile plot function that takes as input \texttt{out}, the output of one of the prediction methods in the package, \texttt{y0}, a list composed by the true values of the response at \texttt{x0}, \texttt{ylab}, a string containing the label for the y-axis, \texttt{titles}, the title for the plot, \texttt{date}, a vector of date strings (for the date formal look at section \nameref{sec:exam}), \texttt{ylim}, the vector of limits for the y-axis and \texttt{fillc}, the fill color for the prediction region. With the help of the packages \pkg{ggplot2} and \CRANpkg{ggpubr}, we can plot all the $q$ dimensions of the response in a single page.

\begin{example}
plot_fun=function(out,y0=NULL,ylab=NULL,titles=NULL,date=NULL,ylim=NULL,fillc="red")
\end{example}

\section{Example}
\label{sec:exam}

As a case study, we examined the BikeMi data, a mobility dataset that tracks all bike rentals for the BikeMi service active in the city of Milan. As in \cite{BikeMi}, we analyzed 41 days, from the 25\textsuperscript{th} of January 2016 to the 6\textsuperscript{th} of March 2016 (excluding the 25\textsuperscript{th} of February), and all data consisted of time and locations of departures and arrivals of bikes. For simplicity, we are considering only the flow from Duomo district to itself as the functional response, and temperature, humidity, and date are used as regressors. 

As the dataset was originally designed for functional modelling, when restricting to the multivariate case some adjustments were required: as response, we used the total number of departures and arrivals from Duomo for every day, while the covariates were the daily averages of the original regressors. Thus, we obtained 41 observations.
In short the model is the following:

\begin{equation}
    \begin{split}
         y_i^k= \beta_0^k + \beta_{we}^kx_{we,i}+\beta_{rain}^kx_{rain,i}+\beta_{temp}^kx_{dtemp,i}+ \beta_{we\_rain}^kx_{we,i}x_{rain,i}+ \epsilon_{i}^k \\   k=1,2 \; \; i=1,...,41
    \end{split}
    \label{eq:overallmodel}
\end{equation}
where k=1 represents the trips starting from Duomo, while k=2 the trips ending in Duomo.
$x_{we,i}$ is a logical equal to 1 if day i is weekend.
$x_{rain,i}$ is the amount of rain (in mm) on day i at time t.
$x_{dtemp,i}$ is the difference from the average daily temperature of the period. 
$x_{we,i}x_{rain,i}$ is an interaction term between weekend and rain. In practice, we can simply retrieve the data, integrated into the package, choose a test point (March the 6\textsuperscript{th}), build the necessary structures, and pick a linear model as regression method with the following code:

\begin{example}
library(conformalInference.multi)
data("bikeMi")
yb<-bikeMi[,1:2]
xb<-bikeMi[,-c(1:2)]
n<-nrow(yb)
y<-as.matrix(yb[1:(n-1),],nrow=(n-1))
x<-as.matrix(xb[1:(n-1),],nrow=(n-1))
y0<-as.matrix(yb[n,],nrow=1)
x0<-as.matrix(xb[n,],nrow=1)
fun=lm_multi()
\end{example}

\begin{figure}[h]
    \centering
    \includegraphics[width=0.85\textwidth]{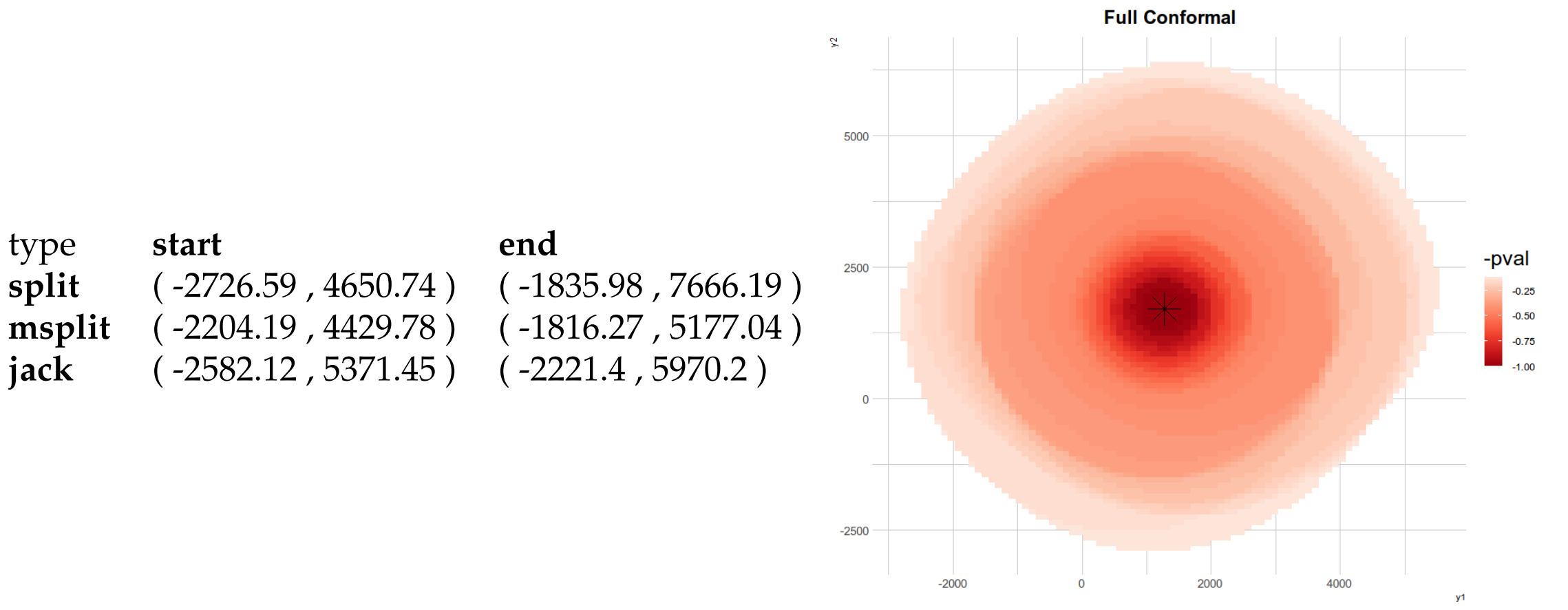}
    \caption{On the left the prediction regions for BikeMi dataset on the 6\textsuperscript{th} of March 2016, while on the right the heatmap of multivariate full conformal prediction region.}
    \label{fig:over_multi}
\end{figure}

For prediction we employed all the methods presented in subsection \nameref{sec:multipre}, with their default values. 
The only exceptions were \texttt{num.grid.pts.dim}, equal to 300 for visualization purposes, the seed in \texttt{conformal.multidim.split} and the modulation function chosen (\textit{alpha-max}). All the methods should yield 90\% percent valid prediction regions. Additionally, in Figure \ref{fig:over_multi} we plotted the full conformal prediction region as well as the produced prediction regions. 

\begin{example}
full<-conformal.multidim.full(x,y,x0, fun$train, fun$predict,num.grid.pts.dim = 300)
plot_multidim(full)
first<-conformal.multidim.split(x,y,x0, fun$train, fun$predict,seed=123,s.type = "alpha-max")
second<-conformal.multidim.msplit(x,y,x0, fun$train, fun$predict)
third<-conformal.multidim.jackplus(x,y,x0, fun$train, fun$predict)

\end{example}

To determine whether the methods are effective, we extracted each observation $(x_0,y_0):=(x_i,y_i) \; i=1,...,n$ from the training set, constructed the prediction set at $x_0$, and tested whether the extracted value $y_0$ was included in the prediction set.  We then averaged the coverage, the mean area, and the computation times across all 41 prediction regions: the results are shown in Table \ref{tab:itermulti}. 

Specifically, one may observe that split conformal method tends to produce wider prediction regions, despite being extremely fast. In contrast, the full conformal method generates extremely accurate predictions, but is once again extremely slow compared to the competition. Ultimately, we found the jackknife+ and multi split extensions to be well-behaved, with reasonable computation times and wide enough regions, while still achieving the 90\% coverage target.

\begin{table}[]
\centering
\begin{tabular}{llll}
type            & \textbf{coverage} & \textbf{avg area} & \textbf{avg time (s)} \\
\textbf{full} & 0.90 &  $5.68*10^7$ & 35.3 \\ 
\textbf{split}  & 0.98              &  $3.46*10^8$       & 0.01      \\
\textbf{msplit} & 0.90              & $4.95*10^7$    & 1.80         \\
\textbf{jack}   & 0.93              & $6.56*10^7$        & 1.96     
\end{tabular}
\caption{Mean coverage, region are and computation time for the multivariate case with BikeMi data.}
\label{tab:itermulti}
\end{table}

In the functional case we considered the original model defined in \cite{BikeMi}.
Here there are two important differences with respect to \eqref{eq:overallmodel}: the regressors and the responses are time-dependent and the bivariate response is the logarithm of the number of trips started ($y^{1}$) and ended ($y^{2}$) in the Duomo district at time t.
The evaluation grid has 90 time steps and is based on a time window between 7.00 A.M. and 1.00 A.M. The model is as follows:

\begin{equation}
    \begin{split}
         log(y_i^k(t))= \beta_0^k(t) + \beta_{we}^k(t)x_{we,i}(t)+\beta_{rain}^k(t)x_{rain,i}(t)+\beta_{temp}^k(t)x_{dtemp,i}(t)+  \\ \beta_{we\_rain}^k(t)x_{we,i}(t)x_{rain,i}(t) +  \epsilon_{i}^k(t)    \; \; k=1,2 \; \; i=1,...,41
    \end{split}
\end{equation}

As the data is already available within the package, we can simply import it.
We divide the data into train and test (for test, just consider day 41), build the evaluation grid for $y$ with \texttt{t\_y} and construct an hour vector for visualization. Finally we select as regression method the concurrent model.

\begin{example}
library(conformalInference.fd)
data("bike_log")
data("bike_regressors")
x<-bike_regressors[1:40]
x0<-bike_regressors[41]
y<-bike_log[1:40]
y0<-bike_log[41]
t_y<-list(1:length(yb[[1]][[1]]),1:length(yb[[1]][[1]]))
hour_test<-seq(from=as.POSIXct("2016-03-06 07:01:30", tz="CET"),
               to=as.POSIXct("2016-03-07 00:58:30", tz="CET"), length.out=90)
fun<-concurrent()
\end{example}

The next step is the actual computation of the bands. Note that for the split conformal method we selected as seed 1234568, allowing repeatability. With the multi split conformal and jackknife+ methods, we chose as number of replication 180 and as $\tau$ the value 0.2.

\begin{example}
first<-conformal.fun.split(x,t_x=NULL,y,t_y,x0,fun$train,fun$predict,seed=1234568)
second<-conformal.fun.msplit(x,t_x=NULL,y,t_y,x0,fun$train,fun$predict,B=180,
tau=0.2)
third<-conformal.fun.jackplus(x,t_x=NULL,y,t_y,x0,fun$train,fun$predict)
\end{example}

\begin{figure}[h]
    \centering
    \includegraphics[width=\textwidth]{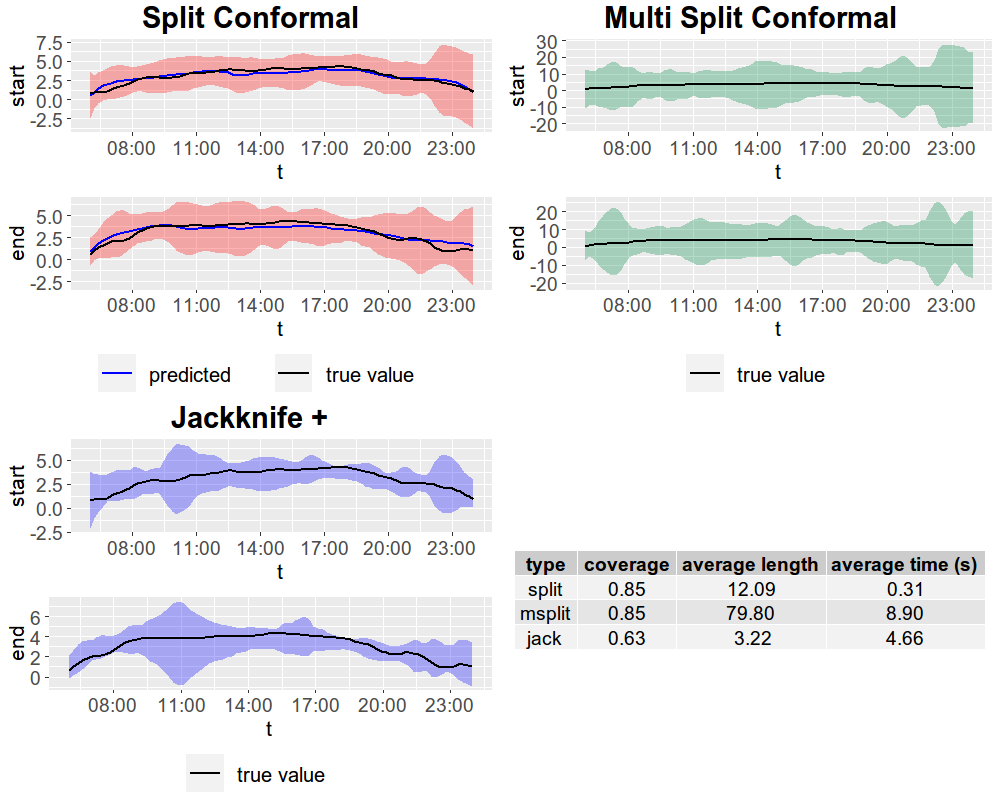}
    \caption{Output of \texttt{plot\_fun} for the three prediction methods. On the bottom right there are the coverage, average length and computation times of the prediction bands.}
    \label{fig:over_fd}
\end{figure}

As a way to display the results we used the \texttt{plot\_fun} function, where we provided the y-axes labels, the titles, the hour vector, the true response at $x_0$, and the fill color as input. Figure \ref{fig:over_fd} shows the plots that were produced.

\begin{example}
plot_fun(first,ylab=c("start","end"),titles="Split Conformal",
         date=list(hour_test,hour_test),y0=y0)
plot_fun(second,ylab=c("start","end"),titles="Multi Split Conformal",
         date=list(hour_test,hour_test),y0=y0,fill="springgreen4")
plot_fun(third,ylab=c("start","end"),titles="Jackknife +",
         date=list(hour_test,hour_test),y0=y0,fill="blue")
\end{example}

The final step in evaluating the predictions was to extract one observation from the training set iteratively and use it as a test value. Subsequently, we averaged the coverage and the mean length of the bands and the computation times. The results are shown in Figure \ref{fig:over_fd}. The multi split conformal method appears to construct wider bands than its simpler counterpart (split conformal), while jackknife+ does not seem as well suited to the analysis, providing extremely compact prediction regions and sacrificing coverage. Overall, the split conformal method is the most suitable for this study, since it provides adequate coverage at a minimal computation time and returns reasonable sized bands. However, the split procedure is subject to randomness.

\section{Summary}

The article recaps the main concepts of Conformal Prediction theory and proposes extensions to the multivariate and functional frameworks of multi split conformal and jackknife+ prediction methods.
Next, the structure and the main functions of \pkg{conformalInference.multi} and \pkg{conformalInference.fd} are extensively discussed. The last section of the paper presents a case study using mobility data collected by BikeMi and compares the prediction methods on the basis of three factors: the size of the prediction regions, the computation time, and the empirical coverage. \\ Therefore, we have bridged the gap between R and other programming languages through the introduction of conformal inference tools for regression in multivariate and functional contexts, and as a result shed light on how versatile as well as effective these methods can be. \\
We envision two future directions for our work.
To begin with, in accordance with its author, we would like to submit \pkg{conformalInference} for publication on CRAN, which would enrich the pool of distribution-free prediction methods available to R users. Secondarily, we would be interested in expanding on the work presented in \cite{timeseries}, to explore conformal inference prediction tools in time series analysis.


\nocite{ggplot2}
\nocite{glmnet}
\nocite{future.apply}
\nocite{future}
\nocite{gridExtra}
\nocite{hrbrthemes}
\nocite{ggpubr}

\bibliography{diquigiovanni-fontana-solari-vantini-vergottini}

  \address{Paolo Vergottini\\
  Mathematical Engineering
  \\Politecnico di Milano\\
  Piazza Leonardo da Vinci, 32, 20133 Milano\\
  Italy\\
 \email{paolo.vergottini@mail.polimi.it}}

\address{Matteo Fontana\\
European Commission, Joint Research Centre \\
  Ispra (VA)\\
  Italy}

\address{Jacopo Diquigiovanni\\
  Department of Statistical Sciences, University of Padova, Italy \\ 
  now at Credit Suisse, Model Risk Management Team, Zurich, Switzerland}

\address{Aldo Solari\\
  Department of Economics, Management and Statistics
  \\University of Milano-Bicocca \\
  Piazza dell'Ateneo Nuovo 1, 20126 Milano\\
  Italy\\
  ORCiD: 0000-0003-1243-0385}

\address{Simone Vantini\\
  MOX-Department of Mathematics
  \\Politecnico di Milano\\
  Piazza Leonardo da Vinci, 32, 20133 Milano\\
  Italy\\
  ORCiD: 0000-0001-8255-5306}

\end{article}

\end{document}